\documentclass[journal]{IEEEtran}
\usepackage[T1]{fontenc}
\usepackage[utf8]{inputenc}
\usepackage{microtype}
\usepackage{booktabs}
\usepackage{graphicx}
\usepackage{amsmath,amssymb}
\usepackage{array}
\usepackage{tikz}
\usetikzlibrary{arrows.meta,positioning}
\usepackage{url}
\usepackage[hidelinks]{hyperref}
\usepackage[caption=false,font=normalsize,
labelfont=sf,textfont=sf]{subfig}
\usepackage{textcomp}
\usepackage{stfloats}
\usepackage{verbatim}
\usepackage{balance}

\title{Neurodiversity-Aware Multimodal Affective Computing for Neurodevelopmental Assessment: From Norm-Referenced Classification to Context-Sensitive Decision Support}

\author{Mateusz Pomianek and Anna \L{}\k{e}\'zniak-Seruga%
\thanks{This work has been submitted to the IEEE for possible publication. Copyright may be transferred without notice, after which this version may no longer be accessible.}

\thanks{M. Pomianek is with the Department of Computer and Control Engineering, Rzeszow University of Technology, Rzesz\'ow, Poland (e-mail: m.pomianek@prz.edu.pl; ORCID: 0000-0001-7769-8919).}%
\thanks{A. \L{}\k{e}\'zniak-Seruga is with the School of Human Sciences, VIZJA University, Warsaw, Poland (e-mail: leznian6857\_aeh@students.vizja.pl; ORCID: 0009-0004-7087-005X).}}

\begin{document}
\maketitle

\begin{abstract}
Automated neurodevelopmental assessment increasingly combines computer vision, speech, eye tracking, physiology, and machine learning, yet multimodality and discrimination do not establish construct validity, clinical usefulness, or appropriate interpretation of behavioral variation. We propose a testable architecture for neurodiversity-aware multimodal affective computing in which population-relative deviation is not treated as sufficient evidence of adverse functioning. This conceptual article reports neither a new participant-level study nor a trained predictive system, and no predictive or clinical superiority is claimed. The framework keeps observable signals, modality quality and availability, context, population-relative information, pooled within-person references and, when sufficiently supported, context-conditioned within-person references, and predictive uncertainty distinguishable throughout inference. It defines structured decision-support outputs and auditable requirements for target validity, temporal integrity, contextual leakage, missing-modality robustness, calibration, subgroup evaluation, interpretability, privacy, and human decision boundaries. The contribution is architectural rather than algorithmic: it specifies empirically testable constraints that can be instantiated with different multimodal learning methods. Autism provides the principal motivating evidence base, without assuming unchanged transfer to other neurodevelopmental conditions.
\end{abstract}


\section{Introduction}

Neurodevelopmental conditions comprise heterogeneous developmental profiles affecting cognition, communication, attention, motor control, executive functioning, social interaction, and adaptive behavior. Their assessment often integrates developmental history, structured screening or diagnostic instruments, and professional observation, motivating computer-assisted methods based on behavioral, physiological, linguistic, electronic-health-record, and neuroimaging data \cite{RajagopalanTammimies2024,Shahini2025,Pang2026}. Prospective digital behavioral phenotyping has also been evaluated as a~scalable adjunct to autism screening in pediatric settings \cite{Perochon2023}.

Affective computing is relevant because clinically informative observations can include dynamic social and affect-related behavior such as facial movement, gaze allocation, head and body motion, vocal prosody, interpersonal timing, and responses to changing task demands. Foundational reviews of affective computing and multimodal affect detection emphasize both the potential value of combining complementary channels and the substantial dependence of performance on task, population, and fusion design \cite{Poria2017,DMelloKory2015,Baltrusaitis2019}. Mobile and video-based studies show that quantifiable behavioral signals can be collected outside specialist laboratories \cite{Egger2018,Perochon2023}, while recent work extends this direction toward multimodal screening and clinical-video understanding using audiovisual or higher-level representations \cite{Natraj2024,Kommineni2024,Pang2026}. These measurements are not treated here as direct readouts of internal emotion or diagnostic state; they are observable signals whose interpretation depends on task, context, data quality, and the inferential target. The framework is specifically an affective-computing contribution because facial dynamics, gaze allocation, vocal prosody, interpersonal timing, arousal-related physiology, and behavioral engagement invite latent affective and social interpretation. The central methodological problem is therefore not only multimodal prediction, but preservation of the inferential boundary between observable affect-related behavior and claims about internal state or clinical meaning. The same observable behavioral signal can acquire different inferential meaning depending on measurement quality, its position in the population, the person's prior trajectory, and the context in which it occurs; these information sources should therefore remain distinguishable until they are combined for an explicitly defined inferential and decision-support purpose.

Technical progress nevertheless exposes a~methodological limitation. Classification performance does not establish that a~model captures clinically meaningful, transferable, or equitable structure. Reviews of machine-learning approaches to neurodevelopmental assessment identify recurring problems involving sample heterogeneity, validation design, generalizability, and translation to real clinical populations \cite{RajagopalanTammimies2024,Shahini2025,Pang2026}. More general work on clinical and psychiatric AI similarly emphasizes that deployment requires attention to external validity, workflow, uncertainty, bias, and human oversight rather than predictive performance alone \cite{Winter2021,Topol2019,Wiens2019}. More fundamentally, a~model optimized to distinguish a~diagnosed group from a~neurotypical comparison group can make distance from the comparison distribution an implicit prediction signal. If that distance is interpreted normatively, variation in gaze, affect-related behavior, communication, movement, or self-regulation can be reduced to deviation from a~population prototype without sufficient representation of within-person variation or environmental context.

The neurodiversity paradigm challenges deficit-only framings of autism research and emphasizes the relevance of person--environment relations \cite{PellicanoDenHouting2022}. Integrative autism-biomarker work likewise argues against overly reductionist single-marker interpretations when dynamic interactions among brain, body, and social environment are central to the phenomenon under study \cite{Loth2023}. Evidence that language-model representations can encode negative associations with neurodivergence-related terms further illustrates that computational bias may operate at the level of representation as well as downstream error rates \cite{Brandsen2024}. A~central premise of the present framework is therefore that statistical deviation from a~population norm is not, by itself, evidence of pathology. Human neurodevelopment does not have a~single normative behavioral phenotype against which every individual can be meaningfully ranked. A~behavioral difference, reduced performance in a~particular domain, or apparent deficit becomes clinically informative only in relation to its functional consequences, persistence, developmental history, environmental demands, compensatory or regulatory role, and the context in which it occurs. Accordingly, not every measured deficit should be treated computationally as a~symptom, and absence of an overtly atypical behavior should not be treated automatically as evidence of unimpaired functioning. This distinction makes contextual information part of the inferential problem rather than an optional covariate.

We therefore define \emph{neurodiversity-aware multimodal affective computing} as computational modeling of observable affect-related and social behavior in which (i) population-relative variation is not assigned adverse clinical meaning independently of context or function, (ii) within-person variation is represented separately from population variation, (iii) modality quality and missingness remain explicit, (iv) inferential uncertainty is preserved at the decision interface, and (v) automated outputs are constrained to a~prespecified screening, monitoring, assessment, or decision-support role. Here, \emph{neurodiversity-aware} denotes an engineering stance requiring explicit treatment of individual variability, context, uncertainty, and person--environment interactions. It neither denies the existence or clinical relevance of disability and impairment nor proposes replacing established diagnostic systems; rather, it constrains how computational observations are converted into inferential claims. A~system should not be considered neurodiversity-aware solely because it uses personalized models, multimodal inputs, or fairness metrics. Under the present framework, the designation additionally requires that population-relative deviation remain distinguishable from adverse functional meaning, that decision-relevant context and uncertainty remain explicit, and that automated outputs remain within a~prespecified human decision pathway.

The contribution is a~technical framework rather than a~new diagnostic classifier. Its novelty does not lie in introducing a~new multimodal fusion algorithm, calibration method, uncertainty estimator, or personalization technique in isolation. Rather, it lies in specifying how established technical components must be jointly constrained when the inferential target concerns heterogeneous neurodevelopmental behavior. In particular, population-relative information, within-person change, context, measurement quality, modality availability, and uncertainty must remain distinguishable and auditable throughout inference so that population deviation cannot silently become equivalent to adverse clinical meaning. Accordingly, the contribution is integrative and architectural rather than algorithmic: it defines a~testable design contract specifying both the conditions under which a~multimodal affective-computing system can be described as neurodiversity-aware and the verification procedures through which those conditions can be empirically challenged.

Specifically, the article: (1) defines a~quality-aware multimodal representation that keeps signal, context, longitudinal history, population-relative information, and individual history as distinguishable information sources, including explicit baseline support and an optional context-conditioned personal reference when repeated comparable observations are available; (2) separates population normalization, subgroup-wise calibration assessment, and within-person modeling as complementary operations answering different statistical questions; (3) defines a~structured decision-support output that preserves uncertainty, modality availability, contextual qualifiers, and traceable model-relevant observations; and (4) converts these principles into auditable design requirements covering target and label validity, temporal integrity, context admissibility, multimodal robustness, calibration, subgroup evaluation, interpretability, privacy, and human decision boundaries. The framework is intended to generate testable engineering hypotheses for affective-computing systems; it does not assert clinical superiority before direct comparative and external validation. Accordingly, the contribution should be evaluated in terms of the internal coherence, operational specificity, and empirical testability of its design requirements rather than the predictive performance of a~particular implementation.

Although the framework is formulated for neurodevelopmental applications broadly, much of the currently available multimodal evidence comes from autism research. Autism is therefore used here as the principal motivating case rather than as evidence that all empirical findings transfer unchanged to ADHD, intellectual disability, developmental language disorder, tic disorders, or other neurodevelopmental conditions. Condition-specific validation remains necessary.

The framework has three explicit non-goals. First, it does not propose replacing categorical diagnosis with individual-baseline modeling. Second, it does not assume that personalization, multimodality, or subgroup calibration is intrinsically fair or clinically beneficial. Third, it does not treat contextual variables as causally informative merely because they improve prediction. Each proposition requires independent empirical evaluation.

Importantly, the proposed framework does not assume that the frequency or intensity of an observable neurodivergent behavior is monotonically related to clinical impairment. The same behavioral manifestation may have different functional significance across individuals and contexts. For example, repetitive self-stimulatory behavior can serve a~self-regulatory function for some autistic people \cite{Kapp2019}, whereas suppression or camouflaging may reduce the observable expression of autistic characteristics without necessarily indicating improved well-being \cite{Zhuang2023}. Consequently, neurodiversity-aware affective computing should model behavior jointly with its individual and situational context rather than treating deviation from a~neurotypical behavioral norm as an independent marker of pathology.

\section{Multimodal neurodevelopmental assessment and screening}

\subsection{From questionnaires to computational behavioral phenotyping}

Automated approaches to autism spectrum disorder (ASD), attention-deficit/hyperactivity disorder (ADHD), and other neurodevelopmental conditions increasingly analyze measurements that are difficult to quantify consistently during routine observation. Video can capture facial dynamics, gaze direction, head orientation, posture, gestures, repetitive movements, and interpersonal timing. Audio permits analysis of speech timing, vocal prosody, rhythm, and acoustic characteristics. Eye tracking can quantify fixations, saccades, scan paths, and pupil-related measures. Physiological sensors and neurophysiological recordings provide additional representations when their acquisition burden is justified. Reviews of current neurodevelopmental AI systems describe this broad expansion from single structured variables toward heterogeneous behavioral and multimodal data \cite{Shahini2025,Pang2026}. The broader multimodal-learning literature likewise treats representation, alignment, fusion, and co-learning as distinct technical problems rather than a~single operation \cite{Baltrusaitis2019}.

Each modality has distinct measurement constraints. Computer-vision pipelines are affected by image quality, occlusion, head pose, illumination, and the validity of the behavioral construct inferred from visual features \cite{DeBelen2020}. Eye-tracking studies depend on calibration, stimulus design, sampling and preprocessing choices, and missing gaze observations \cite{JeyaraniSenthilkumar2023,Iwauchi2023}. Speech-based approaches are affected by language, developmental stage, recording conditions, and task context, while multimodal systems additionally require synchronization and explicit handling of modality-specific missingness \cite{Pang2026}. These limitations are reasons to model data quality explicitly rather than evidence against multimodal analysis itself.

Recent ASD research illustrates the transition from isolated features toward temporally organized behavioral representations. Prospective digital-phenotyping work has combined gaze, facial dynamics, head movement, response to name, and motor measures collected through a~tablet-based application \cite{Perochon2023}. Wearable and eye-tracking studies have quantified gaze patterns during socially relevant tasks \cite{Nag2020,Iwauchi2023}. Video-based visual-attention estimation has also been evaluated for automated ASD and symptom-severity prediction in preschool children \cite{DeBelen2024}. Privacy-preserving datasets increasingly represent behavior through skeletons, body meshes, and optical flow rather than identifiable raw video \cite{Ravva2024}. Multimodal screening studies combine video and audio information \cite{Natraj2024}, while clinical-video work has explored long-form representations that connect speech, visual descriptions, and language-model reasoning \cite{Kommineni2024}. These developments motivate treating the behavioral episode, rather than an isolated frame or single feature, as a~useful computational unit.

\subsection{Limits of reported classification accuracy}

High diagnostic or screening performance has been reported in selected datasets and controlled or prospectively collected cohorts \cite{Perochon2023,Natraj2024,Shahini2025,Pang2026}. Such results demonstrate technical promise but do not establish that neurodevelopmental diagnosis is computationally solved. Reviews of AI-based assessment repeatedly identify sensitivity to cohort composition, label definition, dataset size, validation design, and external generalization \cite{RajagopalanTammimies2024,Shahini2025,Pang2026}. Participant-level separation between training and test data is especially important when multiple samples or recordings originate from the same child, because sample-level splitting can otherwise overstate generalization.

For screening systems, accuracy alone is insufficient. Sensitivity, specificity, area under the receiver-operating-characteristic curve, precision-recall behavior, positive and negative predictive values, calibration, and uncertainty should be interpreted according to intended use. Calibration and predictive uncertainty are distinct technical properties: modern neural networks can be poorly calibrated even when discriminative performance is strong, and uncertainty estimates may degrade under dataset shift \cite{Guo2017,Ovadia2019}. The prospective SenseToKnow study illustrates why this distinction matters: despite high sensitivity, specificity, and area under the curve, positive predictive value was substantially lower than negative predictive value in the screened clinical sample \cite{Perochon2023}. A~balanced case-control benchmark therefore addresses a~different statistical problem from population screening, where prevalence and referral pathways influence predictive value.

\subsection{Affective and behavioral modalities}

Facial analysis can quantify expression dynamics, facial movement, head pose, engagement-related behavior, and temporal responses to social cues \cite{DeBelen2020,Egger2018,Perochon2023,DeBelen2024}. In developmental assessment, timing and change may be more informative than a~static predicted emotion. Higher-level labels should therefore remain distinguishable from directly observed movement. In particular, brief facial movement should not be treated as a~direct readout of an internal emotional state unless the inference has been independently validated for the population and task under study.

Eye tracking offers a~quantitative view of visual attention and social orienting. Fixation duration, gaze transitions, scan paths, and related measurements can reveal group and individual differences in information sampling \cite{Nag2020,Iwauchi2023,JeyaraniSenthilkumar2023}. However, an atypical gaze pattern is not by itself evidence of reduced social interest or deficient processing. Its interpretation depends on stimulus, task, sensory conditions, communication demands, and the relation between gaze behavior and observed functioning. This is one reason single-marker biomarker approaches have limited interpretive scope in heterogeneous neurodevelopmental populations \cite{Loth2023}.

Vocal rate, pitch variability, prosodic timing, pause structure, and conversational coordination can characterize communication behavior, while head and body movement can capture motor strategies, interaction timing, or repetitive action. Multimodal affect-detection research has repeatedly shown that fusion can add information, but also that gains depend strongly on the naturalness of the task and the strength of the unimodal baselines \cite{DMelloKory2015,Poria2017}. Multimodal ASD studies likewise demonstrate that combining complementary behavioral channels can outperform or enrich individual modalities, but the gain is empirical rather than guaranteed \cite{Natraj2024,Pang2026}. Physiological signals can similarly characterize regulatory or arousal-related dynamics without being specific to one diagnostic category. The broader methodological principle is therefore to treat these measurements as context-dependent behavioral or physiological markers rather than automatic deficit markers. Analogous precision-neuropsychiatry work emphasizes the value of combining complementary physiological evidence rather than interpreting a~single measure as a~stand-alone diagnostic proxy \cite{Kovacevic2025}. Because these systems concern children, data minimization, consent, explainability, and developmentally appropriate human oversight also require explicit treatment as design constraints \cite{Chng2025}.

Taken together, these developments shift the principal methodological question from whether neurodevelopmentally relevant behavior can be measured computationally to how such measurements should be interpreted. More modalities can increase observable information without resolving the reference against which variation should be evaluated, the context in which it acquires meaning, or the uncertainty with which an inference should be acted upon. These unresolved questions motivate the neurodiversity-aware framework developed in the following section.

\section{A neurodiversity-aware conceptual framework}

\subsection{Normative bias and the modeling target} %3A

Most supervised diagnostic models are trained on group labels. This is statistically convenient but conceptually restrictive. When a~model is optimized to separate a~diagnosed group from a~population comparison group, features correlated with group membership can become strong prediction signals. We use the term \emph{normative bias} in a~specific computational sense. It does not denote the mere use of population-level reference data, case--control labels, or~discriminative classification. Rather, it denotes a~modeling or interpretive failure in which population-relative deviation is treated, explicitly or implicitly, as sufficient evidence of adverse functioning without establishing the contextual, functional, developmental, or decision-relevant meaning of that deviation. Thus, a~classifier can use population information without exhibiting normative bias, provided that deviation from the reference distribution is not itself assigned a~fixed adverse clinical meaning. No claim is made that case--control classification or population normalization is intrinsically inappropriate. Neurodiversity scholarship has argued for moving beyond deficit-only framings and for treating person--environment relations as scientifically relevant rather than peripheral \cite{PellicanoDenHouting2022,Loth2023}.

Normative bias is not identical to demographic bias. A~system can reduce disparities in error rates across demographic groups while still defining success primarily as conformity to a~neurotypical reference. Conversely, personalization does not guarantee demographic fairness. Bias analyses in mental-health and clinical AI show that sampling, labeling, feature construction, model development, and deployment can all introduce systematic inequities \cite{Timmons2023,Rajkomar2018}. Bias can also enter learned representations themselves: language-model studies have documented negative associations with neurodivergence-related terms even when the downstream task is not explicitly diagnostic \cite{Brandsen2024}. A~neurodiversity-aware system should therefore evaluate both the assumptions embedded in the modeling target and the distribution of errors across populations.

\subsection{Individual baselines and longitudinal variation} %3B

A central proposal of this framework is to compare observations not only with population distributions but also with the individual's own profile over time. This is a~design recommendation rather than a~claim that individual-baseline models have already been established as superior across neurodevelopmental assessment or screening tasks. Its rationale follows from the substantial heterogeneity documented in autism and from the limitations of reductionist single-biomarker approaches \cite{Loth2023,Pang2026}. Repeated within-person measurements can, in principle, distinguish stable idiosyncratic behavior from change associated with context or developmental time. Importantly, an individual baseline is descriptive rather than normative: it summarizes previously observed behavior and does not denote the person's healthy, optimal, typical, or clinically preferred state. It may itself change with development, environment, support, treatment, medication, learning, or other longitudinal influences.

This principle naturally supports longitudinal assessment. Instead of asking only whether an observation resembles a~diagnostic group, a~model can ask whether regulation, participation, communication, gaze allocation, movement, or response to support changes across repeated measurements. As depicted in Figure~\ref{fig:reference-architecture} population and individual representations need not be mutually exclusive; a~hierarchical system can retain population-level information while representing within-person deviation separately. A~further distinction becomes necessary when the individual's prior observations span materially different environments or tasks. A~pooled personal history asks whether the current observation differs from the person's overall prior trajectory, whereas a~context-conditioned history asks whether it differs from that trajectory under prespecified comparable conditions. The latter requires repeated within-context observations and therefore introduces a~support--specificity trade-off. Population-relative, person-relative, and person-within-context representations are complementary rather than progressively more valid: they answer different statistical questions, and none independently establishes clinical abnormality. Section~IV formalizes both personal-reference alternatives and their leakage-safe fallback behavior. The empirical value of such representations must be tested prospectively rather than assumed.

Three statistical operations should not be conflated: population normalization, subgroup-wise calibration assessment, and longitudinal personal-reference modeling. Population normalization describes where an observation lies relative to a~reference distribution. Subgroup-wise calibration assessment asks whether a~probabilistic model output retains its empirical meaning within prespecified groups; it does not by itself impose or guarantee fairness. Longitudinal personal-reference modeling instead compares the current observation with information available previously for the same person, using either pooled history or, when sufficiently supported, a~prespecified context-conditioned subset. These operations answer different questions and are complementary rather than interchangeable.

\subsection{Context as part of the representation} %3C

Behavior occurs under conditions. For the present framework, context is separated into at least five analytically distinguishable classes where relevant: task context, environmental context, interpersonal context, support or intervention context, and developmental or temporal context. Task structure, social demand, sensory stimulation, noise, novelty, fatigue, adult support, peer presence, and transition demands can alter what is observed. The claim here is methodological: context should not automatically be treated as nuisance variance when the research question concerns behavioral adaptation or participation. This position is consistent with neurodiversity-oriented accounts that emphasize interactions between the individual and environment \cite{PellicanoDenHouting2022} and with integrative biomarker perspectives that explicitly include dynamic social and environmental mechanisms \cite{Loth2023}.

We therefore propose representing behavioral episodes as linked components: antecedent or contextual conditions, directly observable behavior, consequences or responses, support strategies, and subsequent observable effects. This structure separates observation from interpretation. Probable behavioral function should remain an observer or model hypothesis rather than be silently promoted to ground truth. One possible operationalization is an antecedent/behavior/consequence/support/effect representation (A/B/C/S/E), in which directly observed events remain distinguishable from hypotheses concerning behavioral function. This representation is illustrative and is not claimed to constitute a~universally validated ontology.

\subsection{Participant representation and data acquisition} %3D

Neurodevelopmental datasets should represent clinically and behaviorally heterogeneous participants rather than relying only on sharply separated case--control groups. Reviews of machine-learning research in autism and other neurodevelopmental conditions note that limited sample diversity and cohort-specific design can undermine generalization \cite{RajagopalanTammimies2024,Shahini2025,Pang2026}. Recruitment and characterization should therefore document variables such as age, sex, language, relevant clinical characteristics, co-occurring conditions, and environmental factors when ethically and scientifically justified. Diagnostic information remains important, but it need not be the sole organizing axis when the research question concerns behavior and context.

Multimodal measurements should be time aligned whenever temporal relationships are analyzed. The processing pipeline should retain modality-specific quality indicators, missingness, calibration state, confidence estimates, and usable observation duration. Multimodal reviews identify missing data, synchronization, and heterogeneous signal quality as recurrent technical problems \cite{Pang2026,Baltrusaitis2019}. Missing data should therefore not be silently converted to a~typical state. If a~face is not visible, gaze cannot be estimated, or speech is not transcribable, absence of a~reliable measurement must remain distinguishable from a~measured zero or neutral value.

\subsection{Privacy-aware representation} %3E

Child behavioral data are highly sensitive. Where scientifically appropriate, models can operate on privacy-reduced representations such as facial landmarks, skeletons, optical flow, acoustic features, or structured behavioral descriptions instead of permanently retaining identifiable raw audiovisual material. MMASD+ demonstrates one concrete implementation based on body meshes, skeleton coordinates, and optical flow for privacy-preserving behavior analysis \cite{Ravva2024}. Such transformations reduce direct exposure of identifiable video but do not eliminate re-identification, governance, or secondary-use risks; privacy protection therefore remains a~system-level requirement rather than a~property guaranteed by feature extraction alone. Child-centered medical-AI guidance similarly emphasizes data governance, consent, transparency, and appropriate human oversight as lifecycle concerns rather than properties of one preprocessing step \cite{Chng2025}.

\subsection{Feature representation and uncertainty} %3F

A neurodiversity-aware feature layer should preserve observable modality-specific measurements, temporal dynamics, context, and measurement quality. Population-standardized values may remain useful for technical comparability, while individual-baseline features can represent deviation from the person's own prior distribution. This combined representation is proposed as a~testable modeling strategy, not as an established clinical standard.

Measurement confidence should remain available during fusion. Computer-vision, gaze, speech, and physiological pipelines all have modality-specific failure conditions \cite{DeBelen2020,JeyaraniSenthilkumar2023,Pang2026}. A~low-confidence modality should therefore remain distinguishable from a~high-quality observation, and missingness should be represented transparently rather than hidden by default imputation. More generally, calibration and uncertainty estimation should be evaluated explicitly because confidence scores can be miscalibrated and may become unreliable under distribution shift \cite{Guo2017,Ovadia2019}.

Table~\ref{tab:design-requirements} translates the conceptual framework into auditable system requirements. The requirements are deliberately phrased as properties that can be tested in an implementation rather than as claims that a~particular architecture already satisfies them.

This distinction is consistent with the ICD-11 Clinical Descriptions and Diagnostic Requirements, which retain categorical autism spectrum disorder diagnoses while explicitly recognizing heterogeneity in manifestations across chronological age, verbal and intellectual ability, and disorder severity, and while using specifiers for co-occurring disorder of intellectual development and degree of functional language impairment \cite{WHOICD11CDDR2024}. The proposed architecture is therefore not intended to replace categorical diagnostic classification. Instead, it~introduces a~computational layer between observable behavior and clinical interpretation, allowing identical behavioral signals to acquire different meanings depending on person-specific baselines, environmental demands, antecedents, and subsequent observable effects.

\section{Modeling, Decision Support, and Evaluation}

\subsection{Reference computational architecture}

To make the framework technically testable, we express it as an auditable information flow rather than a prescribed estimator, neural architecture, or training objective. The purpose is to keep quantities with different inferential roles distinguishable: measured signal, modality availability, measurement quality, population-relative information, within-person information, context, longitudinal state, and uncertainty. For participant $i$, time $t$, and modality $m$, let $\mathbf{x}^{(m)}_{i,t}$ be the measurement, $a^{(m)}_{i,t}\in\{0,1\}$ its availability, $\mathbf{q}^{(m)}_{i,t}$ its quality descriptor, $\mathbf{c}_{i,t}$ the decision-time context, and $\mathbf{h}_{i,<t}$ prior history. Depending on modality, quality can include signal-to-noise ratio, tracking confidence, visible-face fraction, transcription confidence, usable duration, synchronization error, or sensor saturation. A modality-specific encoder gives

\begin{equation}
\mathbf{z}^{(m)}_{i,t}=\phi_m(\mathbf{x}^{(m)}_{i,t};\boldsymbol{\theta}_m).
\label{eq:modality-encoder}
\end{equation}

Equation~\eqref{eq:modality-encoder} allows identity mappings, hand-crafted features, probabilistic measurement models, or learned encoders; modalities need not share dimensionality before an explicit common-space mapping. Its purpose is to retain modality provenance before fusion.

Quality and availability remain separate. Availability answers whether a channel can enter inference; quality describes how much trust should be placed in an available measurement. A non-negative measurement-reliability score,

\begin{equation}
\rho^{(m)}_{i,t}=g_m(\mathbf{q}^{(m)}_{i,t},\mathbf{c}_{i,t}),
\qquad \rho^{(m)}_{i,t}\geq0,
\label{eq:reliability}
\end{equation}

must use only measurement-relevant context available at decision time; diagnosis, referral, or service-status proxies must not enter merely because they predict acquisition success. Importantly, $\rho^{(m)}_{i,t}$ is confidence in the measurement channel, not a probability of a behavioral or diagnostic state. Low gaze-tracking confidence, for example, should reduce reliance on gaze rather than be interpreted as ``neutral gaze.'' One transparent fusion baseline is

\begin{equation}
\alpha^{(m)}_{i,t}=
\frac{a^{(m)}_{i,t}\rho^{(m)}_{i,t}}
{\sum_j a^{(j)}_{i,t}\rho^{(j)}_{i,t}},
\qquad
\mathbf{s}_{i,t}=\sum_m\alpha^{(m)}_{i,t}\mathbf{z}^{(m)}_{i,t},
\label{eq:quality-fusion}
\end{equation}

when the denominator is positive; otherwise the system emits an insufficient-information state rather than fabricating a~prediction.  Equation~\eqref{eq:quality-fusion} is only a reference case. For the weighted-sum reference baseline, each modality representation is mapped to a common d-dimensional space; alternative fusion schemes may retain unequal modality dimensions. Concatenation with explicit quality variables, mixture-of-experts gating, attention, Bayesian fusion, or decision-level fusion are compatible if missingness, quality, and modality provenance remain auditable. More complex fusion should therefore demonstrate value beyond a transparent quality-aware baseline rather than being preferred solely for architectural complexity.

% Reference architecture
\begin{figure*}[t]
\centering
\begin{tikzpicture}[
    font=\scriptsize,
    >=Latex,
    box/.style={
        draw,
        rounded
        corners=1.5pt,
        align=center,
        text width=0.145\textwidth,
        minimum height=14mm,
        inner sep=2.5pt},
    wide/.style={
        draw,
        rounded corners=1.5pt,
        align=center,
        text width=0.16\textwidth,
        minimum height=14mm,
        inner sep=2.5pt},
    governance/.style={
        draw,
        rounded corners=1.5pt,
        align=center,
        text width=0.54\textwidth, %68-69
        minimum height=12mm,
        inner sep=3pt
        },
    arrow/.style={-Latex, line width=0.65pt, shorten >=1pt},
    audit/.style={-Latex, dashed, line width=0.55pt, shorten >=1pt}
]

\node[box] (inputs) at (0,0) {
    \textbf{Modality observations}\\[1.5pt]
    $\mathbf{x}^{(m)}_{i,t}$\\
    quality $\mathbf{q}^{(m)}_{i,t}$\\
    availability $a^{(m)}_{i,t}$};

\node[box, right=4mm of inputs] (fusion) {
    \textbf{Quality-aware fusion}\\[2.5pt]
    $\rho^{(m)}_{i,t}$, $\alpha^{(m)}_{i,t}$\\
    $\mathbf{s}_{i,t}$};

\node[wide, right=4mm of fusion] (representation) {
    \textbf{Contextual / longitudinal representation}\\[1.5pt]
    population $\mathbf{z}^{\mathrm{pop}}_{i,t}$\\
    pooled person $\mathbf{d}_{i,t}$\\
    optional person-in-context $\mathbf{d}^{(c)}_{i,t}$\\
    support $\boldsymbol{\beta}_{i,t}$\\
    context $\mathbf{c}_{i,t}$; history $\mathbf{h}_{i,<t}$};

\node[box, right=4mm of representation] (prediction) {
    \textbf{Prediction, calibration, uncertainty}\\[1.5pt]
    $f_{\theta}$, $C(\cdot)$\\
    $\widehat{\mathbf{p}}_{i,t}$, $\mathbf{u}_{i,t}$};

\node[wide, right=4mm of prediction] (output) {
    \textbf{Structured output and action policy}\\[1.5pt]
    $\mathcal{O}_{i,t}=$\\
    $(\widehat{\mathbf{p}},\mathbf{u},\mathbf{q},\mathbf{a},$\\
    $\mathbf{e},\mathbf{c},\boldsymbol{\beta})$\\
    $A_{i,t}=D(\mathcal{O}_{i,t};\boldsymbol{\tau})$};

\draw[arrow] (inputs.east) -- (fusion.west);
\draw[arrow] (fusion.east) -- (representation.west);
\draw[arrow] (representation.east) -- (prediction.west);
\draw[arrow] (prediction.east) -- (output.west);

\node[governance, below=10mm of representation] (constraints) {
    \textbf{Cross-cutting constraints:}
    only information legitimately available at decision time $t$;
    unavailable personal or within-context history is explicit
    rather than silently imputed; target/label and context
    provenance are auditable; calibration, subgroup behavior,
    missing-modality robustness, interpretability, privacy,
    and human decision boundaries are evaluated.};

\draw[audit] (constraints.north west) -- (fusion.south);
\draw[audit] (constraints.north) -- (representation.south);
\draw[audit] (constraints.north east) -- (prediction.south);
\draw[audit] (constraints.east) -| (output.south);

\node[align=center, font=\scriptsize, below=1.5mm of constraints.south]
    {Dashed links denote evaluation and governance constraints, not learned signal paths};

\end{tikzpicture}
\caption{Reference architecture for neurodiversity-aware multimodal affective computing. Modality-specific measurements retain explicit quality and availability; population-relative, pooled person-relative, and supported person-within-context representations remain distinguishable; contextual and historical inputs are restricted to information legitimately available at the intended decision time; predictive uncertainty and baseline-support metadata remain visible at the decision interface. The final action policy is separated from model inference, and unavailable personal or within-context history is represented explicitly rather than silently imputed.}
\label{fig:reference-architecture}
\end{figure*}
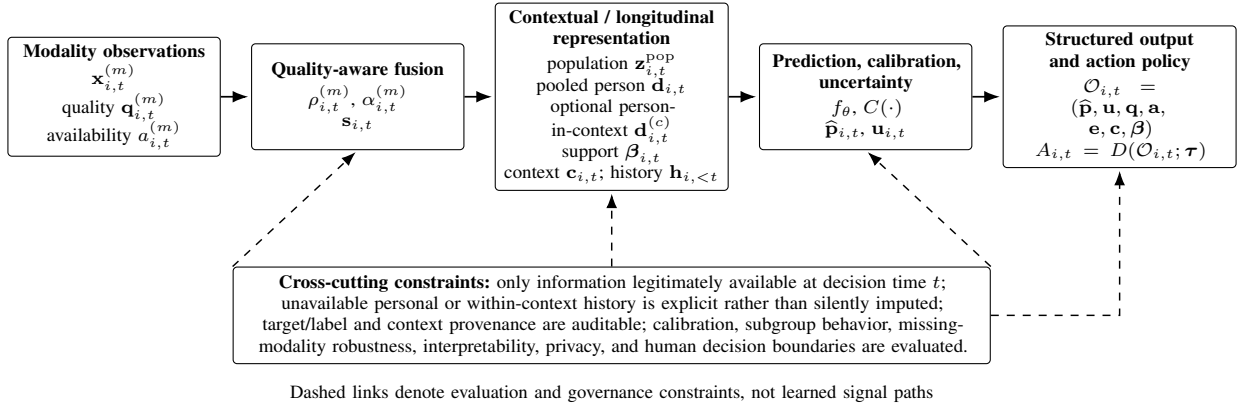

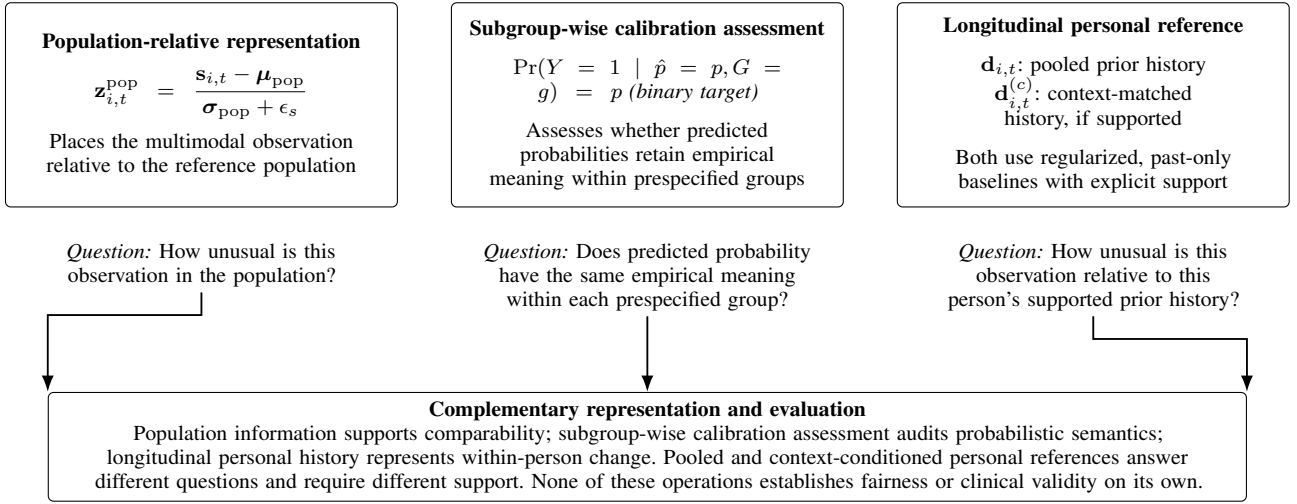
\begin{figure*}[t]
\centering
% Complementary statistical operations for neurodiversity-aware modeling
% Vector source intentionally kept in TikZ for reproducible publication output

\begin{tikzpicture}[
  font=\footnotesize,
  node distance=5mm,
  panel/.style={draw, rounded corners=2pt, align=center, text width=0.27\textwidth, minimum height=27mm, inner sep=4pt},
  note/.style={align=center, text width=0.27\textwidth, inner sep=2pt},
  arrow/.style={->, >=Latex, line width=0.7pt}
]
\node[panel] (pop) {\textbf{Population-relative representation}\\[2mm]
$\mathbf{z}^{\mathrm{pop}}_{i,t}=\dfrac{\mathbf{s}_{i,t}-\boldsymbol{\mu}_{\mathrm{pop}}}{\boldsymbol{\sigma}_{\mathrm{pop}}+\epsilon_s}$\\[2mm]
Places the multimodal observation relative to the reference population};

\node[panel, right=7mm of pop] (sub) {\textbf{Subgroup-wise calibration assessment}\\[2mm]
$\Pr(Y=1\mid \hat p=p,G=g)=p$ \emph{(binary target)}\\[2mm]
Assesses whether predicted probabilities retain empirical meaning within prespecified groups};

\node[panel, right=7mm of sub] (ind) {\textbf{Longitudinal personal reference}\\[2mm]
$\mathbf{d}_{i,t}$: pooled prior history\\
$\mathbf{d}^{(c)}_{i,t}$: context-matched history, if supported\\[2mm]
Both use regularized, past-only baselines with explicit support};

\node[note, below=4mm of pop] (popn) {\textit{Question:} How unusual is this observation in the population?};
\node[note, below=4mm of sub] (subn) {\textit{Question:} Does predicted probability have the same empirical meaning within each prespecified group?};
\node[note, below=4mm of ind] (indn) {\textit{Question:} How unusual is this observation relative to this person's supported prior history?};

\node[draw, rounded corners=2pt, align=center, text width=0.86\textwidth, minimum height=13mm, below=10mm of subn] (joint) {\textbf{Complementary representation and evaluation}\\
Population information supports comparability; subgroup-wise calibration assessment audits probabilistic semantics; longitudinal personal history represents within-person change. Pooled and context-conditioned personal references answer different questions and require different support. None of these operations establishes fairness or clinical validity on its own.};

\draw[arrow] (popn.south) -- ++(0,-3mm) -| (joint.north west);
\draw[arrow] (subn.south) -- (joint.north);
\draw[arrow] (indn.south) -- ++(0,-3mm) -| (joint.north east);
\end{tikzpicture}
\caption{Distinct reference operations: population normalization, subgroup-wise calibration assessment, and pooled or sufficiently supported context-conditioned personal-reference modeling. None is sufficient for fairness or clinical validity.}
\label{fig:reference-models}
\end{figure*}

Figs.~\ref{fig:reference-architecture} and ~\ref{fig:reference-models} summarize the reference architecture and distinguish three commonly conflated operations. Population normalization asks where an observation lies relative to a reference distribution. Subgroup-wise calibration assessment asks whether probabilistic outputs retain their empirical meaning within prespecified groups; it does not guarantee fairness. Personal-reference modeling asks how the current observation differs from information previously available for the same person, either across pooled history or within sufficiently supported comparable contexts. Table~\ref{tab:framework-comparison} summarizes this contrast. Retaining these operations separately allows their incremental value and failure modes to be tested rather than assuming that one reference frame should replace another.

\subsubsection{Population-relative and within-person references}

Let $\boldsymbol{\mu}_{\mathrm{pop}}$ and $\boldsymbol{\sigma}_{\mathrm{pop}}$ denote dimension-matched training/reference-population parameters, with the operations in \eqref{eq:population-deviation}--\eqref{eq:person-deviation} interpreted componentwise; the scalar case is a special case. Population-relative position is
\begin{equation}
\mathbf{z}^{\mathrm{pop}}_{i,t}=
\frac{\mathbf{s}_{i,t}-\boldsymbol{\mu}_{\mathrm{pop}}}
{\boldsymbol{\sigma}_{\mathrm{pop}}+\epsilon_s}.
\label{eq:population-deviation}
\end{equation}

It describes statistical position, not deficit, symptom, severity, diagnosis, or adverse functioning. Reference parameters must be estimated without test leakage, and any stratification or covariate adjustment must be specified as part of the model rather than introduced after inspecting test performance.

For prior-person history, let $n_{i,t}$ be the number of valid observations before $t$, with sample moments $\boldsymbol{\mu}_{i,<t}$ and $\boldsymbol{\sigma}^2_{i,<t}$ when estimable. A simple cold-start regularization is

%\begin{align}
\begin{equation}
\lambda_{i,t}=\frac{n_{i,t}}{n_{i,t}+\kappa},\qquad \kappa>0,
\label{eq:shrinkage-weight}
\end{equation}
with
%\end{align}

\begin{align}
\widetilde{\boldsymbol{\mu}}_{i,<t}
&=\begin{cases}
\boldsymbol{\mu}_{\mathrm{pop}},&n_{i,t}=0,\\
\lambda_{i,t}\boldsymbol{\mu}_{i,<t}
 +(1-\lambda_{i,t})\boldsymbol{\mu}_{\mathrm{pop}},&n_{i,t}\ge1,
\end{cases}
\label{eq:shrinkage-mean}\\
\widetilde{\boldsymbol{\sigma}}^2_{i,<t}
&=\begin{cases}
\boldsymbol{\sigma}^2_{\mathrm{pop}},&n_{i,t}<2,\\
\lambda_{i,t}\boldsymbol{\sigma}^2_{i,<t}
 +(1-\lambda_{i,t})\boldsymbol{\sigma}^2_{\mathrm{pop}},&n_{i,t}\ge2.
\end{cases}
\label{eq:shrinkage-var}
\end{align}

The leakage-safe within-person deviation is

\begin{equation}
\mathbf{d}_{i,t}=
\frac{\mathbf{s}_{i,t}-\widetilde{\boldsymbol{\mu}}_{i,<t}}
{\sqrt{\widetilde{\boldsymbol{\sigma}}^2_{i,<t}+\epsilon_s^2}}.
\label{eq:person-deviation}
\end{equation}

Equations~\eqref{eq:shrinkage-weight}--\eqref{eq:person-deviation} make the cold start explicit. At first observation the baseline is population-supported; after one prior observation the location can begin to personalize while the scale remains population-supported; repeated observations progressively increase the individualized contribution. Numerical shrinkage must not be confused with existence of a reliable personal baseline. The parameter $\kappa$ is not universal and should be prespecified or estimated within training data and examined in sensitivity analysis. Robust estimators, Bayesian hierarchical models, state-space models, Gaussian processes, exponentially weighted baselines, or learned temporal encoders are compatible with the framework; the invariant constraints are forward-only information flow, explicit support, and separation of population- and person-relative information. A personal baseline remains descriptive rather than normative.

A support flag can prevent shrinkage from creating the appearance of a mature personal baseline:

\begin{equation}
\beta_{i,t}=\mathbb{I}[n_{i,t}\ge n_{\min}],
\label{eq:baseline-support}
\end{equation}

where $n_{\min}$ and $\kappa$ are prespecified or learned only within training data and subjected to sensitivity analysis. If $\beta_{i,t}=0$, the individualized component can be masked or reported with a low-support flag.

A synthetic contrast makes the semantics concrete. Suppose a scalar feature has the same current value for two participants, $s_{A,t}=s_{B,t}=0.50$, but supported prior-person baselines $(\widetilde{\mu},\widetilde{\sigma})=(0.44,0.02)$ and $(0.52,0.02)$. Equation~\eqref{eq:person-deviation} gives approximately $d_{A,t}=3.0$ and $d_{B,t}=-1.0$. With $\mu_{\mathrm{pop}}=0.40$ and $\sigma_{\mathrm{pop}}=0.10$, both observations have $z^{\mathrm{pop}}\approx1.0$. Identical measurements can therefore have different within-person meanings. Large $|d_{i,t}|$ means unusual relative to supported prior history, not pathological, improved, or deteriorated. The example is illustrative, not empirical validation.

Context has two roles that should remain distinct. Current context can condition interpretation or prediction, whereas recurrent context can determine which prior observations form an appropriate within-person comparison set. The first changes inference; the second changes the reference against which longitudinal change is quantified.

When recurrent contexts are sufficiently sampled, the historical comparison set may be restricted prospectively to

$\mathcal{T}^{(c)}_{i,t}=\{\tau<t:\Gamma(\mathbf{c}_{i,\tau},\mathbf{c}_{i,t})=1\}$,

where $\Gamma$ defines admissible contextual comparability. The same shrinkage logic can then estimate a~context-conditioned personal reference, with explicit within-context support and fallback to the pooled personal baseline. If support is insufficient, context and $\mathbf{d}_{i,t}$ remain separate inputs. Context matching must preserve temporal causality and exclude variables downstream of diagnosis, referral, clinician knowledge, or service allocation unless those variables are explicitly part of the intended decision pathway. The definition of $\Gamma$ should be fixed before model evaluation and its support reported. Population-relative, person-relative, and person-within-context representations are complementary, not interchangeable estimators of clinical abnormality.

\subsubsection{Context, prediction, uncertainty, and action}

Context may condition inference but can also create shortcuts if it encodes downstream consequences of prior diagnosis, referral, or service allocation. Only variables legitimately available at the intended decision time are admissible. A contextual representation can be written as

\begin{equation}
\mathbf{r}_{i,t}=
\psi(\mathbf{s}_{i,t},\mathbf{z}^{\mathrm{pop}}_{i,t},
\mathbf{d}_{i,t},\mathbf{c}_{i,t},\mathbf{h}_{i,<t},
\mathbf{q}_{i,t},\mathbf{a}_{i,t}),
\label{eq:contextual-representation}
\end{equation}

optionally substituting or adding a supported context-conditioned deviation. History and personal references must use information strictly before $t$.

A task-specific predictor maps $\mathbf{r}_{i,t}$ to raw scores and then, when probabilistic output is required, to a predictive distribution $\widehat{\mathbf{p}}_{i,t}$. Calibration must be estimated on training/validation data only and evaluated in the intended population and prespecified subgroups rather than inferred from discrimination \cite{Guo2017}. A simple categorical uncertainty measure is normalized predictive entropy,

\begin{equation}
u^{\mathrm{pred}}_{i,t}
=-\frac{1}{\log K}\sum_{k=1}^{K}
\widehat p_{i,t,k}\log(\widehat p_{i,t,k}+\epsilon_p),
\label{eq:predictive-entropy}
\end{equation}

but predictive entropy alone is insufficient. Uncertainty can arise from ambiguous observations, limited model knowledge, degraded measurements, missing modalities, weak personal-baseline support, or distribution shift \cite{Ovadia2019}. A system-level descriptor $\mathbf{u}_{i,t}$ can therefore combine predictive uncertainty with optional epistemic/OOD indicators and the quality, availability, and baseline-support metadata. The framework does not prescribe one estimator; it requires uncertainty and insufficient information not to disappear behind a single precise-looking score. The final interface is therefore structured:

\begin{equation}
\mathcal{O}_{i,t}=
(\widehat{\mathbf{p}}_{i,t},\mathbf{u}_{i,t},
\mathbf{q}_{i,t},\mathbf{a}_{i,t},
\mathbf{e}_{i,t},\mathbf{c}_{i,t},\boldsymbol{\beta}_{i,t}),
\label{eq:output-tuple}
\end{equation}

where $\mathbf{e}_{i,t}$ contains traceable model-relevant observations or attributions linked to provenance, not independent clinical evidence. For screening, $\widehat{\mathbf{p}}_{i,t}$ may represent calibrated referral risk; for contextual tasks it may represent a behavioral-state distribution or probability of an observable response under specified conditions. The tuple is a decision-support interface, not a diagnostic conclusion. A separate policy maps inference to action,

\begin{equation}
A_{i,t}=D(\mathcal{O}_{i,t};\boldsymbol{\tau}),
\label{eq:decision-rule}
\end{equation}

with prespecified thresholds and abstention criteria. Low quality, unsupported personalization, or high uncertainty can trigger repeat acquisition rather than referral or reassurance. Thresholds depend on intended use, prevalence, availability of confirmatory assessment, and asymmetric consequences of false-negative and false-positive decisions. The policy therefore separates statistical inference from the clinical or operational pathway in which it is used.

\begin{table*}[t]
\caption{Conventional classification versus the proposed framework}
\label{tab:framework-comparison}
\centering
\footnotesize
\setlength{\tabcolsep}{3pt}
\begin{tabular}{p{0.14\textwidth}p{0.26\textwidth}p{0.53\textwidth}}
\toprule
\textbf{Dimension} & \textbf{Conventional classifier} & \textbf{Proposed framework} \\
\midrule
Target & Diagnostic or case--control label & Prespecified behavioral, contextual, longitudinal, functional, or referral-related target \\
Reference & Population comparison & Population reference retained separately from supported personal references \\
Context & Omitted or nuisance covariate & Explicit when scientifically relevant and decision-time admissible \\
Missing/quality & Often handled in preprocessing & Availability and quality retained through inference and audit \\
Time & Optional aggregation & Causal longitudinal history and baseline support explicit \\
Output & Class label or score & Calibrated prediction plus uncertainty, quality, context, support, and provenance \\
Evaluation & Primarily discrimination & Discrimination, calibration, generalization, subgroup audit, ablation, and robustness \\
Clinical role & Risk of conflation with diagnosis & Screening, assessment, monitoring, or support kept distinct from diagnosis \\
\bottomrule
\end{tabular}
\end{table*}

Table~\ref{tab:design-requirements} turns the computational chain into a compact verification contract.

\begin{table*}[t]
\caption{Auditable requirements for neurodiversity-aware multimodal affective computing}
\label{tab:design-requirements}
\centering
\scriptsize
\setlength{\tabcolsep}{3pt}
\begin{tabular}{p{0.16\textwidth}p{0.28\textwidth}p{0.27\textwidth}p{0.23\textwidth}}
\toprule
\textbf{Requirement} & \textbf{Design principle} & \textbf{Minimum verification} & \textbf{Failure signal} \\
\midrule
Target validity & Prespecify target, label provenance, and intended screening/assessment/monitoring role & Intended-use statement; label provenance; reliability or sensitivity analyses where applicable & Historical label or referral practice is treated as an~objective person property \\
Context and leakage & Preserve relevant task/environment/interpersonal /support context, but admit only decision-time variables & Temporal provenance; context ablation; proxy/shortcut analysis & Person-level inference is driven by setting, prior diagnosis, or service-status proxies \\
Within-person reference & Keep population and personal features distinct; prespecify pooled versus supported context-conditioned baseline; baselines are descriptive & Leakage-safe temporal split; component ablation; pooled/context-conditioned comparison with support reporting & Future leakage, unstable sparse baselines, or contextual shifts misread as person-level change \\
Cold start & Make insufficient personal or within-context history explicit & First-contact versus mature-baseline analysis; report support flags; sensitivity to support parameters & Missing history is silently treated as typical or as target evidence \\
Quality-aware multimodality & Preserve modality availability, quality, synchronization, and usable duration & Missing-modality and degradation tests; modality provenance & Missing/poor input is interpreted as a~neutral state \\
Temporal integrity & Respect observation order in baselines and history & Forward-only construction and participant-level temporal validation & Future observations influence current inference \\
Uncertainty-aware output & Couple predictions with uncertainty, quality, and support metadata & Calibration; uncertainty/OOD or low-quality flags where feasible & Precise output despite insufficient information \\
Subgroup evaluation & Audit calibration and errors in prespecified groups, including intersections when supported & Group metrics with uncertainty and sample sizes & Aggregate performance conceals subgroup failure or sparse cells are overinterpreted \\
Interpretability & Link explanations to observable inputs, provenance, context, and quality; avoid mechanistic overclaiming & Ablation, perturbation, and explanation-stability checks & Explanations persist after relevant inputs are removed or corrupted \\
Governance and human role & Minimize identifiable data and keep output within a defined professional decision pathway & Data-flow/access review; consent/governance audit; documented escalation boundaries & Privacy-reduced features are treated as risk-free or output becomes autonomous diagnosis \\
\bottomrule
\end{tabular}
\end{table*}

\subsection{Evaluation strategy}

No fusion strategy is intrinsically neurodiversity-aware. Early fusion can model cross-modal dependencies but requires stronger alignment and can increase dimensionality; intermediate fusion preserves modality-specific encoders before interaction; late fusion can simplify operation when channels are unavailable. Attention-based or mixture-based alternatives may also be appropriate. Regardless of architecture, quality and provenance should remain accessible for missing-modality tests and ablation \cite{Baltrusaitis2019,Pang2026}. Multimodal studies show that fusion can improve selected unimodal baselines, but the gain is task-dependent rather than guaranteed \cite{Natraj2024,Huynh2025}. Sequential models are similarly justified when transitions, persistence, timing, or context-dependent change are part of the scientific question, provided that temporal causality is preserved. Long-form clinical-video and script-centered work illustrate emerging temporal representations without establishing their universal validity across neurodevelopmental tasks \cite{Kommineni2024,Liu2026}.

Evaluation should address at least five complementary dimensions. Predictive performance includes task-appropriate discrimination, predictive values, uncertainty intervals, and prespecified decision thresholds. Calibration measures whether predicted probabilities retain their empirical meaning. Generalization tests performance across sites, devices, demographic groups, and independent cohorts. Fairness analysis examines subgroup-wise calibration and error behavior, including intersectional groups when adequately supported. Interpretability and robustness require modality ablation, missing-modality tests, quality degradation, contextual perturbation, and checks that explanations track observable inputs rather than artifacts. These dimensions address different failure modes and should not be collapsed into headline accuracy. Subgroup calibration is not a complete definition of fairness, and sparse intersectional cells require sample-size and uncertainty reporting \cite{Timmons2023,Rajkomar2018}.

The formulation yields direct nested ablations: population-only versus pooled personal versus sufficiently supported context-conditioned personal references; unweighted versus quality-aware fusion; context-free versus contextual prediction; raw versus calibrated outputs; and class-score-only versus structured uncertainty-aware decision support. The more complex variant should not be assumed superior. Ablation is particularly important in multimodal systems because it reveals whether apparent gains come from complementary information or disproportionate dependence on one channel. Screening outputs should likewise be evaluated within their operational pathway: screening is not diagnosis, predictive value depends on prevalence and referral consequences, and a useful output may combine calibrated risk with quality, uncertainty, and contextual qualifiers rather than expose only a class score \cite{Perochon2023,RajagopalanTammimies2024}. Human-factor evaluation is therefore needed to determine whether additional information improves decisions rather than merely increasing interface complexity.

\subsection{Ethical and clinical governance}

False negatives and false positives have asymmetric consequences that depend on the care pathway, so decision thresholds cannot be justified by discrimination alone. Screening is not diagnosis: a system may identify cases warranting further assessment without replacing comprehensive clinical evaluation \cite{Perochon2023,Shahini2025}. Interpretability is relevant to accountability, but post-hoc explanation is not evidence of a learned clinical mechanism; explanation stability should be checked against input perturbation and ablation \cite{Ghassemi2021}. Users should be able to determine which modalities were available, what quality limitations were present, how uncertain the output is, and whether the observation lies outside the evaluated distribution. Privacy requires minimization, secure processing, consent, access control, retention rules, and purpose limitation even when identifiable media are transformed into reduced representations \cite{Ravva2024}. Neurodivergent people and caregivers should be involved in defining beneficial outcomes, acceptable data use, and meaningful explanations \cite{PellicanoDenHouting2022}.

\section{Discussion}

The framework addresses a~specific gap in computational neurodevelopmental assessment: stronger sensing and classification do not resolve construct validity, transferability, calibration, or interpretation of behavioral variation \cite{RajagopalanTammimies2024,Shahini2025,Pang2026}. Its central constraint is a~staged inferential boundary: a~sensor produces an observation rather than a~construct; a~population-relative score describes statistical position rather than pathology; a within-person deviation describes change rather than improvement or deterioration; context conditions interpretation without establishing causality; and a calibrated prediction remains an input to, not a replacement for, a clinical or operational decision.

The proposal complements rather than replaces established diagnostic or dimensional frameworks. RDoC emphasizes functional constructs across units of analysis and recognizes developmental and environmental influences \cite{NIMHRDoC2026,CuthbertInsel2013}; biopsychosocial accounts likewise place functioning within interactions among biological, psychological, and social factors \cite{Engel1977}. DSM-5-TR and ICD-11 retain categorical diagnoses while representing clinically meaningful heterogeneity through specifiers and functional characterization \cite{DSM5TR2022,WHOICD11CDDR2024}. The present contribution operates at a~different computational level: it specifies how multimodal observations, quality metadata, context, leakage-safe personal history, uncertainty, and decision-support outputs should remain distinguishable and auditable. Its novelty lies in this operational and verification architecture, not in proposing a~new taxonomy of psychopathology.

Informal feedback from clinicians and psychologists with neurodevelopmental-care experience was used to refine the intended role and boundary conditions of the framework. It was not collected or analyzed as a~formal qualitative study and is not empirical validation. The feedback emphasized that longitudinal assessment extends beyond assigning a~category at one time point: clinicians observe functioning, response to support or treatment, and change as a person's needs evolve. This is particularly relevant when several neurodevelopmental, psychiatric, or other clinical conditions coexist, because an observed change may not map uniquely to one diagnosis. Accordingly, the architecture is designed to preserve person-specific and contextual observations for professional interpretation rather than prescribe treatment or replace clinical judgment.

Population normalization, subgroup-wise calibration assessment, pooled personal baselines, and supported context-conditioned baselines answer different statistical questions. Population statistics provide a~reference distribution; calibration asks whether predicted probabilities retain their meaning; pooled personal history represents change relative to the person's overall trajectory; and context-conditioned history compares observations under prespecified similar conditions. Comparable subgroup calibration is not a~complete fairness criterion, and personalization is not a~fairness guarantee. Individual models can retain sampling, measurement, or demographic bias, while subgroup estimates can become unstable in sparse data. Likewise, multimodal sensing is useful only when modality-specific quality, alignment, and missingness remain visible; otherwise additional channels can introduce new error sources rather than reliable evidence \cite{Guo2017,Ovadia2019,Baltrusaitis2019}.

Multimodal sensing nevertheless remains important because different channels characterize complementary aspects of the same behavioral episode. Prospective digital phenotyping and multimodal screening demonstrate the feasibility of combining gaze, facial dynamics, movement, audio, and structured behavioral information \cite{Perochon2023,Natraj2024,Huynh2025}. The broader affective-computing literature likewise supports fusion while showing that gains depend on task naturalness, modality quality, alignment, and the strength of unimodal baselines \cite{DMelloKory2015,Poria2017}. The framework therefore does not argue against multimodal classification; it requires that gains remain attributable to observable inputs and that synchronization error, missingness, and degraded measurements remain visible in the inference chain.

The same discipline applies to interpretation. Facial movement, gaze, vocal behavior, movement, and physiological signals are not direct proxies for diagnostic status or internal emotion. Their evidential meaning can differ across individuals and contexts, and a~behavioral pattern that is population-atypical may be adaptive, compensatory, regulatory, or clinically neutral in a~particular setting. This is consistent with heterogeneous autism phenotypes and with biomarker perspectives that emphasize dynamic interactions among person, body, and environment \cite{PellicanoDenHouting2022,Loth2023}. The practical value of the architecture is therefore not that it resolves clinical meaning automatically, but that it preserves the information needed to qualify that meaning.

\subsection{Limitations and boundary conditions}

Individual-baseline modeling requires repeated observations and may be unavailable at first contact. Context can be difficult to standardize across naturalistic settings, can encode site-specific or downstream proxies, and can create sparse person--context combinations. A~pooled personal baseline may confound genuine longitudinal change with recurrent changes in task, environment, interpersonal conditions, or support, whereas a~context-conditioned baseline can reduce that ambiguity only at the cost of additional sparsity and estimation variance. Its comparability rules and fallback behavior therefore need prospective definition. Subgroup calibration and error metrics are unstable in small or unrepresentative samples; representation bias may escape standard error audits \cite{Timmons2023,Brandsen2024}; explanation methods can be unfaithful at the individual level \cite{Ghassemi2021}; and privacy-reduced features can remove useful information without eliminating governance or re-identification risk \cite{Ravva2024}. Higher-level generated descriptions introduce an additional verification problem because semantic summaries may fail to preserve clinically relevant audiovisual evidence faithfully \cite{Kommineni2024}.

These limitations define empirical tests rather than implementation details. The framework should be challenged through nested comparisons of population-only, pooled-person, and supported context-conditioned references; context-free and contextual models; complete- and missing-modality operation; and class-score-only versus structured uncertainty-aware outputs. Evaluation should include calibration, decision-threshold behavior, subgroup performance, degraded-input robustness, label and context provenance, and leakage-safe longitudinal validation. External validation is necessary because context definitions, sensor characteristics, referral pathways, and population composition can differ substantially between sites. The framework also requires prospective definition of admissible context, temporal ordering, baseline-update rules, and contextual-comparability criteria so that apparently improved personalization cannot be produced by future leakage or post-hoc subgroup construction. If a richer representation adds variance, depends on shortcuts, or fails to generalize, the simpler model should be preferred. Contextualization and personalization are therefore hypotheses to test, not assumed benefits.

\section{Conclusion}

Neurodiversity-aware affective computing requires more than adding modalities, personalization, or fairness metrics to a diagnostic classifier. The proposed architecture keeps observable signal, population-relative position, within-person change, context, measurement quality, uncertainty, and decision action analytically distinct. Its design claims are falsifiable through ablation, calibration, temporal validation, subgroup audit, missing-modality analysis, and external evaluation.

The framework would be challenged if personal references add no value over population-only models on relevant longitudinal tasks, context-conditioned baselines add variance without benefit, contextual models rely on diagnosis or service-status shortcuts, structured uncertainty fails to improve decision quality, or new subgroup disparities emerge. These failure conditions are important because neurodiversity awareness is treated here as a set of testable engineering constraints rather than an assumed property of any personalized model. This article therefore establishes a~formal architectural basis rather than evidence of clinical superiority. A companion empirical study in preparation investigates one operationalization of context-sensitive behavioral representation using observational data and is not presented as validation of the complete multimodal architecture.

\section*{Author contributions}
M. Pomianek: conceptualization, methodology, framework design, literature verification, visualization, writing---original draft, and writing---review and editing.
A. \L{}\k{e}\'zniak-Seruga: conceptualization and writing---review and editing.

\section*{Data and code availability}
No new participant-level data, executable software, or trained predictive models are reported in this conceptual article. Auditability is supported through explicit representation definitions, temporal constraints, cold-start rules, structured outputs, and prespecified verification criteria. A companion empirical ContextNDD study is in preparation to investigate one operationalization of context-sensitive behavioral representation; it is not presented as validation of the complete architecture.

\section*{Acknowledgment}
During manuscript preparation, OpenAI ChatGPT-5.6 Sol was used for language editing, LaTeX-code assistance, and bibliography-consistency checking. The scientific hypotheses, conceptual framework, modeling assumptions, evaluation criteria, interpretation, and conclusions were determined by the authors. AI-assisted output was subject to human review and correction before inclusion in the manuscript. AI-generated content was not used as scientific evidence for manuscript claims. AI systems do not qualify for authorship. Responsibility for scientific accuracy, integrity, and interpretation remains with the human authors.

\bibliographystyle{IEEEtran}
\bibliography{references,references_review2}

\end{document}